\documentstyle[preprint,aps]{revtex}
\input epsf.tex
\def\DESepsf(#1 width #2){\epsfxsize=#2 \epsfbox{#1}}
\begin{document}
\preprint{\vbox{\hbox{}}}
\draft
\title{CP Violating $b\to s \gamma$ Decay in Supersymmetric Models}
\author{$^1$Chun-Khiang Chua, $^{1,2}$Xiao-Gang He, and $^{1,3}$Wei-Shu Hou}
\address{
$^1$ Physics Department, National Taiwan University, 
Taipei, Taiwan 10764, R.O.C.\\
$^2$School of Physics, University of Melbourne,
Parkville, Vic. 3052, Australia\\
\rm $^3$ Physics Department, Brookhaven National Laboratory,
Upton, NY 11973, USA
}
\date{August, 1998}
\maketitle
\begin{abstract}
Supersymmetric models with nonuniversal squark masses can enrich
the chiral structure and CP violating phenomena in $b\to s\gamma$ decays.
Direct CP violation in $b\to s \gamma$ decay, 
mixing induced CP violation in radiative $B_{d,s}$ decays 
(such as $B_s\to \phi \gamma$ and $B_d \to K^*_{1,2}\gamma$), and 
$\Lambda$ polarization in $\Lambda_b \to \Lambda \gamma$ decay
can be substantially different from the Standard Model. 
Future experiments at $e^+ e^-$ and hadronic B factories will give important
information on the underlying couplings for radiative $b$ decays.
\end{abstract}


\preprint{\vbox{\hbox{}}}

\preprint{\vbox{\hbox{}}}

\preprint{\vbox{\hbox{}}}


\section{Introduction}

The processes $B\to K^{*}\gamma $ and $b\to s\gamma $ are the 
first penguin processes to be observed in B decays \cite{1st expt}. 
As quantum loop effects, they provide good tests for 
the Standard Model (SM).
The measured branching ratios\cite{new expt} are 
in agreement with SM predictions\cite{Chetyrkin,Kagan}, 
although new physics effects are still allowed \cite{HeMc}. 
To further test SM, one must study the detailed structure of 
the $b s\gamma $ couplings. 
In SM, the quark level $b s\gamma $ coupling is 
usually parametrized as 
\begin{equation}
H_{\rm SM}=-c_{7}^{{\rm SM}}\,{\frac{G_{F}}{\sqrt{2}}}
       {\frac{e}{8\pi ^{2}}} V_{tb}V_{ts}^{*}
       \bar{s}[m_{b} (1+\gamma _{5})+m_{s}(1-\gamma _{5})]
       \sigma_{\mu \nu }F^{\mu \nu }b,  \label{ }
\end{equation}
where $c_{7}^{{\rm SM}} \cong -0.3$ at 
the typical B decay energy scale $\mu \approx 5$ GeV. 
One notable feature is that 
the $1+\gamma _{5}$ chiral structure dominates, 
which reflects the left-handed nature of weak interactions. 
Although the branching ratio measurements are consistent with SM, 
they can not determine the chiral structure of the couplings. 
In models beyond SM, it is in principle possible that 
both chiralities are comparable, and 
the $1-\gamma _{5}$ component may even be the dominant one. 
It is therefore important to experimentally confirm 
the chiral structure.

The chirality structure can be tested by studying CP violation. 
If the $1+\gamma _{5}$ chiral structure dominates completely, 
then only direct CP violating rate asymmetries are possible. 
Since such asymmetries are small in SM \cite{Soares}, 
their observation would 
indicate the presence of new physics. 
If both chiralities are present, possible only with new physics,
mixing induced CP violation can occur as well\cite{Atwood}. 
Independent of CP violation, the chirality structure can also 
manifest itself in radiative beauty baryon decays, leading to 
different polarizations of the final state baryon\cite{Mannel}. 
When these asymmetries and polarizations are measured,
they will provide useful information on the 
underlying couplings for radiative b decays.

In this paper we show that in supersymmetric models with 
nonuniversal squark mass matrices, the chiral structure for 
$b\to s\gamma$ can be very different from SM. 
We then illustrate how the chiral structures can be
studied by direct and mixing induced CP violation 
as well as $\Lambda$ polarization in 
inclusive $b\to s\gamma$, exclusive $B\to M \gamma$, 
and $\Lambda_b \to \Lambda \gamma$ decays,
respectively.

\section{Radiative $B$ Decay in Supersymmetric Models}

Supersymmetry (SUSY) is one of the 
leading candidates for physics beyond SM \cite{SUSY}. It can 
help resolve many potential problems when one 
extends beyond SM, for example the gauge hierarchy problem, 
unification of SU(3)$\times $SU(2)$\times $U(1) gauge couplings, 
and so on. 
SUSY models also lead to many interesting low energy phenomena. 
We will concentrate on flavor changing $b\rightarrow s\gamma$ 
decay due to nonuniversal squark masses. 

Potentially large new flavor and CP violating sources 
may come from interactions between 
quarks, gauginos, higgsinos and squarks \cite{Bertolini}. 
These interactions are given by
\begin{eqnarray}
L =-\sqrt{2}g_{s} \left(
      \overline{d}_{L} \Gamma_{DL}^{\dagger}
      -\overline{d}_{R} \Gamma _{DR}^{\dagger}
                  \right) T^{a}\widetilde{D}\,\widetilde{g}_{a}
   -g\widetilde{U}_{k}^{*}\overline{\widetilde{\chi}}_{j}^{c}
      \left[\left(G_{UL}^{jki}-H_{UR}^{jki}\right)P_{L}
            -H_{UL}^{jki}P_{R} \right]d_{i}+h.c.,
\end{eqnarray}
where $P_L$ is the left-handed projection, 
$\widetilde {Q}$, $\widetilde{g}$ and $\widetilde{\chi}$ are 
the squark, gluino and chargino fields, and 
$j$, $k$, $i$ are summed from 1 to 2, 1 to 6 and 1 to 3, respectively. 
The $\Gamma_{QL,R}$ matrices are the mixing matrices that relate
the weak eigenstates $\widetilde{Q}_{L,R}^{i}$ to 
the mass eigenstates $\widetilde{Q}^{k}$,
\begin{equation}
(\widetilde{Q}_{L},\ \widetilde{Q}_{R})=(\Gamma _{QL}^{\dagger },\ 
\Gamma_{QR}^{\dagger })\widetilde{Q}.
\end{equation}
The matrices $G$, $H$ are related to $\Gamma_{QL,R}$
and the chargino mixing matrices $U$ and $V$\cite{SUSY} by 
\begin{eqnarray}
G_{UL}^{jki} = V_{j1}^{*}\,(\Gamma_{UL}V_{\rm CKM})^{ki},\ \ \ 
H_{UL}^{jki} = U_{j2}\,(\Gamma_{UL}V_{\rm CKM}\widehat{Y}_{D})^{ki},\ \ \ 
H_{UR}^{jki} = V_{j2}^{*}\,(\Gamma_{UR} \widehat{Y}_{U}V_{\rm CKM})^{ki},
\end{eqnarray}
where $V_{\rm CKM}$ is the CKM quark mixing matrix, 
$\widehat{Y}_{D}= $~diag($m_{d},m_{s},m_{b})/(\sqrt{2} M_{W}\cos\beta)$ 
and 
$\widehat{Y}_{U}= $~diag($m_{u},m_{c},m_{t})/(\sqrt{2} M_{W}\sin\beta)$.
Note that, in contrast to Ref. \cite{Bertolini}, 
we have kept the $V_{\rm CKM}$ factor explicitly
in $G_{UL}$ and $H_{UL}$ rather than absorbing it into
$\Gamma_{UL}$.

Inspired by minimal supergravity models, 
the usual approach to SUSY modelling is to 
assume universal soft SUSY breaking masses. 
This certainly reduces the number of parameters, 
but it also removes soft squark masses as 
a potent source of flavor (and CP) violation. 
As we are concerned with the possible impact of SUSY models 
on $b\rightarrow s\gamma $ decay, 
we consider general low energy mass mixings 
without assuming specific forms for 
the squark mass matrix at high energies. 
There is then no theoretical constraint on 
the form of $\Gamma _{QL,R}$ at the SUSY breaking scale. 

One might expect that the dominant contributions come from 
gluino exchange because the coupling is stronger. 
However, it has been shown that chargino contributions 
can be important if flavor and CP violation in
$\Gamma_{QL}$ are large\cite{ng}. 
We will therefore include these contributions as well. 
There are also contributions from neutralino exchange. 
We have analyzed neutralino contributions and 
find their contributions to be about one order of 
magnitude smaller in the parameter space we consider. 

The effective Hamiltonian due to 
gluino-squark and chargino-squark exchange
for $b\rightarrow s\gamma ,\ sg$ transitions is given by 
\begin{eqnarray}
H_{{\rm eff.}} &=&-{\frac{G_{F}}{\sqrt{2}}}{\frac{e}{8\pi ^{2}}}%
V_{tb}V_{ts}^{*}m_{b}\,\bar{s}\left[ c_{7}\,(1+\gamma _{5})+c_{7}^{\prime
}\,(1-\gamma _{5})\right] \sigma _{\mu \nu }F^{\mu \nu }b  \nonumber \\
&&-{\frac{G_{F}}{\sqrt{2}}}{\frac{g}{8\pi ^{2}}}V_{tb}V_{ts}^{*}m_{b}\,\bar{s%
}\left[ c_{8}\,(1+\gamma _{5})+c_{8}^{\prime }\,(1-\gamma _{5})\right]
\sigma _{\mu \nu }T^{a}G_{a}^{\mu \nu }b,
\end{eqnarray}
where we have neglected $m_{s}$, 
and $c_{7,8}=c_{7,8}^{{\rm SM}}+c_{7,8}^{{\rm %
New}}$ are the sum of SM and new physics contributions, while $%
c_{7,8}^{\prime }$ come purely from new physics. 
They are given by
\begin{eqnarray}
c_{7}^{{\rm New}}(M_{W}) &=&
 \frac{\sqrt{2}\pi \alpha _{s}}{{G}_{F}V_{tb}V_{ts}^{*}}
 \frac{Q_{d}C_{2}(R)}{m_{\widetilde{D}_k}^{2}}
 \left\{(\Gamma _{DL}^{\dagger})^{sk}f_{2}(a_{\widetilde{g}k})
         \Gamma _{DL}^{kb}
        -\frac{m_{\widetilde{g}}}{m_{b}}\,\
        (\Gamma _{DL}^{\dagger})^{sk}f_{4}(a_{\widetilde{g}k})
         \Gamma _{DR}^{kb}
 \right\}  \nonumber\\
&&+\frac{1}{V_{tb}V_{ts}^{*}}\frac{M_{W}^2}{m_{\widetilde U_k}^2}
 \left\{\left(G_{UL}^{jkb}-H_{UR}^{jkb}\right)
        \left(G_{UL}^{jks}-H_{UR}^{jks}\right)^{*}
        \left[f_{1}(b_{jk})+Q_u\,f_{2}(b_{jk})\right]
 \phantom{\frac{m_{\widetilde{\chi }_{j}^{-}}}{m_{b}}}
 \right.
\nonumber\\
&&\hspace{1.18in}-H_{UL}^{jkb}
 \left. \left(G_{UL}^{jks}-H_{UR}^{jks}\right)^{*}
        \frac{m_{\widetilde{\chi }_{j}^{-}}}{m_{b}}
        \left[f_{3}(b_{jk})+Q_u\,f_{4}(b_{jk})\right]
 \right\},
\\
c_{8}^{{\rm New}}(M_{W}) &=&
 \frac{\sqrt{2}\pi \alpha _{s}}{{G}_{F}V_{tb}V_{ts}^{*}}
 \left\{ {\frac{2C_{2}(R)-C_{2}(G)}{2m_{\widetilde{D}_k}^{2}}}
 \left[ (\Gamma _{DL}^{\dagger})^{sk}f_{2}(a_{\widetilde{g}k})
         \Gamma _{DL}^{kb}
       -\frac{m_{\widetilde{g}}}{m_{b}}\,
        (\Gamma _{DL}^{\dagger})^{sk}f_{4}(a_{\widetilde{g}k})
         \Gamma _{DR}^{kb}
 \right]
 \right.   \nonumber \\
&&\hspace{1.4in} \left. -{\frac{C_{2}(G)}{2m_{\widetilde{D}_k}^{2}}}
 \left[ (\Gamma_{DL}^{\dagger })^{sk}f_{1}(a_{\widetilde{g}k})
         \Gamma _{DL}^{kb}
       -\frac{m_{\widetilde{g}}}{m_{b}}\,
        (\Gamma_{DL}^{\dagger })^{sk} f_{3}(a_{\widetilde{g}k})
         \Gamma _{DR}^{kb}\right]
 \right\} 
\label{c8New} \nonumber\\
&&+\frac{1}{V_{tb}V_{ts}^{*}}\frac{M_{W}^2}{m_{\widetilde U_k}^2}
 \left\{\left(G_{UL}^{jkb}-H_{UR}^{jkb}\right)
        \left(G_{UL}^{jks}-H_{UR}^{jks}\right)^{*}\,f_{2}(b_{jk})
 \phantom{\frac{m_{\widetilde{\chi }_{j}^{-}}}{m_{b}}}
 \right.
\nonumber\\
&&\hspace{1.2in}-H_{UL}^{jkb}
 \left. \left(G_{UL}^{jks}-H_{UR}^{jks}\right)^{*}
 \frac{m_{\widetilde{\chi }_{j}^{-}}}{m_{b}}\,f_{4}(b_{jk})
 \right\}.
\end{eqnarray}
where $Q_{d,u}$ are the electric charges of the down and up type quarks,
$a_{\widetilde{g}k}\equiv m_{\widetilde{g}}^2/m_{\widetilde D_k}^2$, 
$b_{jk}\equiv m_{\widetilde{x}_{j}^{-}}^{2}/{m}_{\widetilde U_k}^{2}$,
$C_{2}(G)=N=3$ and $C_{2}(R)=(N^{2}-1)/(2N)=4/3$ are Casimirs,
and the functions $f_{i}(x)$ are given by 
\begin{eqnarray}
f_{1}(x) &=&\frac{1}{12\,(x-1)^{4}}(x^{3}-6x^{2}+3x+2+6x\,\ln x),
 \nonumber\\
f_{2}(x) &=&\frac{1}{12\,(x-1)^{4}}(2x^{3}+3x^{2}-6x+1-6x^{2}\,\ln x),
 \nonumber\\
f_{3}(x) &=&\frac{1}{2\,(x-1)^{3}}(x^{2}-4x+3+2\,\ln x), \nonumber\\
f_{4}(x) &=&\frac{1}{2\,(x-1)^{3}}(x^{2}-1-2x\,\ln x),
\end{eqnarray}
which agree with Ref. \cite{Bertolini}. 
The first term of Eqs. (6) and (7) comes from gluino exchange 
while second term comes from chargino exchange.
The chirality partners $c_{7,8}^{\prime }$ from gluino exchange 
are obtained by interchanging $\Gamma _{QL}$ and $\Gamma _{QR}$.
The chargino contributions to $c'_{7,8}$ are suppressed by $m_s/m_b$.
The $\Gamma _{DL}^{\dagger }(...)\Gamma _{DL}$ terms arise from 
mixing among $\widetilde{D}_{L}$ alone, while 
$\Gamma _{DL}^{\dagger }(...)\Gamma _{DR}$ terms come from 
mixing between $\widetilde{D}_{L}$ and $\widetilde{D}_{R}$.
We denote these as LL and LR mixing, respectively. 
Note the $M_{\widetilde{g}}/m_{b}$ enhancement factor for LR mixing. 
The $m_{\widetilde{\chi }_{j}^{-}}/m_{b}$ 
enhancement factor in chargino contribution is 
softened by a factor of $m_{b}/M_{W}$ in $H_{UL}^{jkb}$. 

When running down to the B decay scale $\mu \approx m_{b}$,
the leading order Wilson coefficients $c_{i}^{(\prime )}$ and 
next to leading order coefficients $c_{7}^{(\prime )(1)}$ 
are given by \cite{Kagan}, 
\begin{eqnarray}
c_{7}(\mu  &=&m_{b})=-0.31+0.67\,c_{7}^{{\rm New}}(M_{W})
                          +0.09\,c_{8}^{{\rm New}}(M_{W}),
 \nonumber\\
c_{8}(\mu  &=&m_{b})=-0.15+0.70\,c_{8}^{{\rm New}}(M_{W}),
 \nonumber \\
c_{7}^{(1)}(\mu  &=&m_{b})=+0.48-2.29\,c_{7}^{{\rm New}}(M_{W})
                                -0.12\,c_{8}^{{\rm New}}(M_{W}),
\end{eqnarray}
while for opposite chirality, which receives no SM contribution, 
one simply replaces $c^{{\rm New}}$ by $c^{\prime }$ 
and set the constant terms to zero.

In obtaining the above expressions we have 
assumed that SUSY breaking occurs at the TeV scale and 
the squark and gluino masses are in the few hundred GeV region. 
Therefore the gluino, squarks, top quark and W boson are 
integrated out at $\mu \approx m_{t}$ at the same time. 
The coefficients obtained can be very different from 
the SM predictions, but they are of course subject to
the constraint from the observed $b\to s\gamma $ branching ratio, 
in the form of 
\begin{eqnarray}
Br(B\rightarrow \left. X_{s}\gamma)\right|_{E_{\gamma }>(1-\delta)\,
                                     E_{\gamma }^{\max }} 
\approx  2.57\times 10^{-3}\times K_{NLO}(\delta)\,\times \,
  \frac{Br(B\rightarrow X_{c}e\overline{\nu })}{10.5\%},
\end{eqnarray}
which should be compared to the most recent 
experimental result of \cite{new expt} $(3.15\pm 0.54)\times 10^{-4}$,
and $\delta $ is a parameter that defines the photon energy cut
(ideally $\delta = 1$). 
We take the last factor in Eq. (10) to be one and 
\begin{equation}
K_{NLO}(\delta )=\sum_{\stackrel{i,j=2,7,8}{i\leq j}}k_{ij}(\delta )\,
\mathop{\rm Re}%
[c_{i}c_{j}^{*}+c_{i}^{\prime }c_{j}^{\prime *}]+k_{77}^{(1)}(\delta )\,
\mathop{\rm Re}%
[c_{7}^{(1)}c_{7}^{*}+c_{7}^{\prime (1)}c_{7}^{\prime *}],
\end{equation}
where $c_{2}^{\prime }=0$ and $k_{ij}(\delta )$ are 
known functions of $\delta$, their values for some $\delta $ 
can be obtained by using the expressions given in Ref. \cite{Kagan}.
We use $\delta =90\%$ which gives 
$Br(B\to X_s \gamma) \approx 3.3\times 10^{-4} $ in SM, 
in good agreement with data.

Because of the large number of parameters in the mixing matrices, 
it is not practical to perform a general analysis 
in the full parameter space. 
Our purpose is to demonstrate that in SUSY models, 
the prediction for CP violation and the
chiral structure can be considerably different from SM predictions. 
We will restrict ourselves to some simple cases, and consider 
{\it mixing only between second and third generation down type squarks}. 
This has the advantage that the usual stringent constraints 
from processes involving the first generation, such as bounds from 
$K^{0}$-$\bar K^0$ mixing, neutron EDM and so on can be evaded easily, 
and hence allow for large CP violation in $B$ decays. 
In general $B_{s}$-$\bar{B}_{s}$ mixing 
would also be different from SM.
Present limits do not impose strong constraint in 
the parameter space we consider, but may become 
more restrictive as experimental bounds improve.

Having decoupled the first generation,
the $4\times 4$ mixing matrix $(\Gamma _{DL},\ \Gamma _{DR})$ 
diagonalizes the squark mass matrix 
\begin{equation}
\widetilde{M}^{2}_{\rm diag} =(\Gamma _{DL},\ \Gamma _{DR})\left( 
\begin{array}{ll}
\widetilde{m}_{LL}^{2} & \widetilde{m}_{LR}^{2} \\ 
\widetilde{m}_{LR}^{2\dagger } & \widetilde{m}_{RR}^{2}
\end{array}
\right) \left( 
\begin{array}{l}
\Gamma _{DL}^{\dagger } \\ 
\Gamma _{DR}^{\dagger }
\end{array}
\right) ,
\end{equation}
and must satisfy the following equations 
\begin{equation}
(\Gamma _{DL}\Gamma _{DL}^{\dagger }
+\Gamma _{DR}\Gamma _{DR}^{\dagger})^{kl}=\delta ^{kl},\ 
\Gamma _{DL(R)}^{\dagger ik}\Gamma_{DL(R)}^{kj}=\delta ^{ij},\ 
\Gamma _{DR(L)}^{\dagger ik}\Gamma_{DL(R)}^{kj}=0,
\end{equation}
where $i,\ j=2$, $3$ and $k,\ l=2,\ 3$, $5,\ 6$. 
We consider some simple cases for illustration:
\begin{description}
\item[(a)] LL or RR mixing: 
Mixing only in $\widetilde{D}_{L}$ sector (LL) 
and/or $\widetilde{D}_{R}$ sector (RR). 

With $\widetilde{m}_{LR}^{2}=0$ while $\widetilde{m}_{LL}^{2}$,
$\widetilde{m}_{RR}^{2}$ are general $2\times 2$ hermitian matrices,
one has $\Gamma _{DL}^{\dagger }=(u^{\dagger },0)$, 
$\Gamma _{DR}^{\dagger}=(0,v^{\dagger })$. 
The unitary matrices $u,$ $v$ satisfy 
$u\,\widetilde{m}_{LL}^{2}u^{\dagger }=
 {\rm diag}\,(\widetilde{m}_{L,2}^{2},\widetilde{m}_{L,3}^{2})$, 
$v\,\widetilde{m}_{RR}^{2}v^{\dagger }=
 {\rm diag}\,(\widetilde{m}_{R,2}^{2},\widetilde{m}_{R,3}^{2})$, 
and 
\begin{equation}
(\Gamma _{DL}^{\dagger })^{sk}\,f(a_{k})\,\Gamma _{DL}^{kb}=
u^{\dagger s2}u^{2b}(f(a_{2})-f(a_{3}))=
\cos \theta \,\sin \theta \,e^{i\sigma}(f(a_{2})-f(a_{3})),
\end{equation}
while 
$(\Gamma _{DL}^{\dagger })^{sk}\,f(a_{k})\,\Gamma _{DR}^{kb}=0$,
with similar relations for $v$. 
There is one mixing angle $\theta$ and one physical phase $\sigma$ 
for both $u$ and $v$. 
Note that the 
{\it phase of $v$ is not constrained} by $B\rightarrow X_{s}\gamma$. 
To further reduce the parameter space we take $\theta$ to be 
the same for $u$ and $v$ but allow the masses to be different. 
There are also two extreme cases of interest: 
LL only i.e. no RR mixing, 
or LL=RR i.e. $\widetilde{m}_{LL}^{2}=$ $\widetilde{m}_{RR}^{2}$.

For simplicity, we take advantage of the fact that
the mass matrices $\widetilde m^2_{U,RR}$ 
and $\widetilde m^2_{U,LR}$ are independent from 
$\widetilde m^2_{RR}$ and $\widetilde m^2_{LR}$,
and assume no $LR$ and $RR$ mixings in the $\widetilde U$ sector.
That is, we take $\widetilde m^2_{U,RR}$ to be diagonal 
and $\widetilde m^2_{U,LR} = 0$.
However, 
since we have LL mixing in $\widetilde{D}$ sector, 
LL mixing in $\widetilde{U}$ sector will follow accordingly 
because of the ${\rm SU}_L(2)$ symmetry of the theory. 
That is,  
\begin{eqnarray}
\widetilde{m}_{LL}^{2} &=& 
 {\rm diag}\, (m_d^2,m_s^2,m_b^2) + M_{\widetilde{Q}}^{^{\prime }2}
 - M_{Z}^{2}\left(\frac{1}{2}+Q_ds_{W}^{2}\right)\cos 2\beta,
\nonumber \\
\widetilde{m}_{U,LL}^{2} &=&
  \text{diag\,(}m^2_{u},m^2_{c},m^2_{t}) +
  V_{\rm CKM}\left[M_{\widetilde{Q}}^{^{\prime}2}
 + M_{Z}^{2}\left(\frac{1}{2}-Q_us_{W}^{2}\right)\cos 2\beta 
             \right] V_{\rm CKM}^{\dagger },
\end{eqnarray}
where $M_{\widetilde{Q}}^{^{\prime }2}$ are soft squark masses.
In our numerical study, we shall illustrate with 
$m^2_{\chi^+_{1,2}} = 200,\;400$ GeV and $\tan \beta = 2$. 
Up type squark masses will 
depend on down type squark masses and mixing angle.
We apply a 100 GeV lower bound on up type squark masses, 
which further constrains the down squark mixing angle.

\item[(b)] LR mixing only. 

We consider an interesting case with 
$\widetilde{m}_{LL}^{2}=\widetilde{m}_{RR}^{2}=
{\rm diag}\,(\widetilde{m}^{2},\ \widetilde{m}^{2})$ and
neglect down-type quark masses,
while $\widetilde{m}_{LR}^{2}$ is a general $2\times 2$ matrix.
In this case, because $M_{\widetilde{Q}}^{^{\prime }2}$ is 
proportional to the unit matrix, 
$\widetilde m^2_{U,LL}$ is diagonal, as can be seen from Eq. (15).
One then sees from Eq. (4) that 
the chargino contributions in Eqs. (7) and (8) 
are proportional to $V_{kb}V_{ks}^*$,
and hence are much smaller than the gluino contributions. 

Diagonalization, 
$\widetilde{M}^{2}_{\rm diag}=
{\rm diag}\,(\widetilde{m}_{2}^{2},\,\widetilde{m}_{3}^{2},\,
\widetilde{m}_{5}^{2},\,\widetilde{m}_{6}^{2})={\rm diag}\,
(\widetilde{m}^{2}+\Delta \widetilde{m}_{2}^{2},\,
\widetilde{m}^{2}+\Delta \widetilde{m}_{3}^{2},\,
\widetilde{m}^{2}-\Delta \widetilde{m}_{2}^{2},\,
\widetilde{m}^{2}-\Delta \widetilde{m}_{3}^{2})$,
is achieved via 
$\Gamma _{L}^{\dagger }=(u^{\dagger },\,
u^{\dagger })/\sqrt{2}$, $\Gamma _{R}^{\dagger }=
(v^{\dagger },\,-v^{\dagger })/\sqrt{2}$, 
where $u$ and $v$ are unitary matrices satisfying 
$u\,\widetilde{m}_{LR}^{2}v^{\dagger }=
{\rm diag}\,(\Delta \widetilde{m}_{1}^{2},\,
\Delta \widetilde{m}_{2}^{2})$. 
One then finds 
\begin{eqnarray}
2\,(\Gamma _{DL}^{\dagger })^{sk}\,f(a_{k})\,\Gamma _{DL}^{kb} 
&=& u^{\dagger s2}u^{2b}(f(a_{2})-f(a_{3})+f(a_{5})-f(a_{6})),
 \nonumber\\
2\,(\Gamma _{DR}^{\dagger })^{sk}\,f(a_{k})\,\Gamma _{DR}^{kb} 
&=& v^{\dagger s2}v^{2b}(f(a_{2})-f(a_{3})+f(a_{5})-f(a_{6})),
 \nonumber\\
2\,(\Gamma _{DL}^{\dagger })^{sk}\,f(a_{k})\,\Gamma _{DR}^{kb}
&=&(u^{\dagger si}f(a_{i})v^{ib}-u^{\dagger si}f(a_{i+3})v^{ib}),
 \nonumber\\
2\,(\Gamma _{DR}^{\dagger })^{sk}\,f(a_{k})\,\Gamma _{DL}^{kb}
&=&(v^{\dagger si}f(a_{i})u^{ib}-v^{\dagger si}f(a_{i+3})u^{ib}),
\end{eqnarray}
where $i$ is summed over 2 and 3, and 
\begin{equation}
u=\left( 
\begin{array}{ll}
\phantom{-}c\,e^{i\tau } & s\,e^{i\sigma } \\ 
-s\,e^{i\tau } & c\,e^{i\sigma }
\end{array}
\right) ,\;v=\left( 
\begin{array}{ll}
\phantom{-}c^{\prime }\,e^{i\tau ^{\prime }}
 & s^{\prime }\,e^{i\sigma^{\prime }} \\ 
-s^{\prime }\,e^{i\tau ^{\prime }}
 & c^{\prime }\,e^{i\sigma ^{\prime }}
\end{array}
\right).
\end{equation}
We further simplify by 
assuming $\widetilde{m}_{LR}^{2}$ to be hermitian, hence $u=v$. 
Since a 2 $\times $ 2 hermitian matrix has 
4 independent real parameters, 2 would lead to 
eigenvalues $\Delta \widetilde{m}_{2}^{2}$ and 
$\Delta \widetilde{m}_{3}^{2}$, and again we have 
just one mixing angle and one phase.
\end{description}

Although these cases are rather simplified, 
they can still lead to phenomenological consequences 
that are very different from SM. 
In the following sections, we proceed to study
i) direct CP violating partial rate asymmetry $A_{{\rm CP}}$, 
ii) mixing induced asymmetry $A_{{\rm mix}}$, and 
iii) final state $\Lambda $ polarization $\alpha_{\Lambda }$ 
in $\Lambda _{b}\to \Lambda \gamma$,
that follow from our model.

\section{Direct CP violation}

The CP violating partial rate asymmetry $A_{\rm CP}$ 
in $b\rightarrow s\gamma$ decay is defined as 
\begin{eqnarray}
A_{{\rm CP}} =
{\frac{\Gamma (b\rightarrow s\gamma )-\bar{\Gamma}(\bar{b}
\rightarrow \bar{s}\gamma )}{\Gamma (b\rightarrow s\gamma )
+\bar{\Gamma}(\bar{b}\rightarrow \bar{s}\gamma )}}
={\frac{|c_{7}|^{2}+|c_{7}^{\prime }|^{2}
 -|\bar{c}_{7}|^{2}-|\bar{c}_{7}^{\prime }|^{2}}{|c_{7}|^{2}
 +|c_{7}^{\prime }|^{2}+|\bar{c}_{7}|^{2}
 +|\bar{c}_{7}^{\prime }|^{2}}},
\end{eqnarray}
where $\bar{c}_{7}^{(\prime )}$ are coefficients for 
$\bar{b}$ decay.
In SM, $A_{\rm CP} \sim 0.5\%$ \cite{Soares} is very small,
it is therefore a good place to look for deviations from SM. 

To have nonzero $A_{{\rm CP}}$, apart from CP violating phases,
one also needs absorptive parts. 
In the model under consideration, the absorptive parts 
come only from the SM contribution with 
$u$ and $c$ quarks in the loop.
Because of the left-handed nature of charged currents in SM,
the absorptive parts in $c'_{2,7,8}$ are suppressed by 
a factor of $m_s/m_b$ which is small 
and therefore can be neglected.
One finds \cite{Kagan} 
\begin{equation}
A_{{\rm CP}}(\delta )=
 \frac{1}{|c_{7}|^{2}+|c_{7}^{^{\prime }}|^{2}}
 \left\{a_{27}(\delta )\;%
\mathop{\rm Im}%
[c_{2}\,c_{7}^{*}]+a_{87}(\delta )\,%
\mathop{\rm Im}%
[c_{8}\,c_{7}^{*}]+a_{28}(\delta )\,%
\mathop{\rm Im}%
[c_{2}\,c_{8}^{*}]\right\} ,
\end{equation}
where the parameters $a_{ij}(\delta )$ depend on $\delta $ 
which defines the photon energy cut 
$E_{\gamma }>(1-\delta )\,E_{\gamma }^{\max }$, as mentioned earlier. 
From Ref. [4], we find that $a_{87}\sim -9.5\%$ is much larger than 
$a_{27}\sim 1.06\%$ and $a_{28}\sim 0.16\%$. Hence 
large $A_{{\rm CP}}$ is likely to occur when $c_{8}$ is sizable. 
We have carried out detailed studies and find that 
there is a large parameter space where $A_{{\rm CP}}$ 
can be substantially larger than the SM prediction. 
We give some special cases from (a) and (b) in Figs. 1 and 2.

Fig. 1 shows $A_{{\rm CP}}$ vs. $\theta $ and $\sigma $ for 
the case with both LL and RR mixings but no LR mixing. 
We take the mass eigenvalues $m_{\tilde{g}}=200$ GeV, 
which we will use in all cases, and 
$(\widetilde{m}_{L,2},\ \widetilde{m}_{L,3},\ \widetilde{m}_{R,2},\ 
\widetilde{m}_{R,3})=(100,\,250,\,100,\,150)$ GeV. 
The asymmetry $A_{{\rm CP}}$ can reach 10\%. 
If we only consider the gluino contribution, 
the allowed region is much reduced and 
$A_{{\rm CP}}$ can only reach a few percent. 
The chargino contribution is important in the sense that it can 
partially cancel against the gluino contribution,
and the allowed region in the parameter space is enlarged.
Naively one might think that the gluino contribution 
dominates over the chargino contribution 
because $\alpha _{\rm s}/\alpha _{\rm w}$ is large. 
However, this factor is only about 3 and is easily overcome 
by other enhancement factors in the chargino sector.
In particular, the function $f_{3}$ in the chargino 
contribution is larger than $f_{1}$ in the gluino contribution.
It turns out that both contributions are about the 
same order of magnitude and partially cancel against each other for 
the parameter space considered.
Thus, $Br(B\rightarrow X_{s}\gamma)$ close to the SM result is 
easier to achieve, hence enlarging the allowed parameter space. 
For large $\tan \beta$, the parameter space is more restrictive 
because the chargino contribution tends to dominate over 
the gluino contribution.
We have checked that the neutralino contribution is 
about one order of magnitude smaller compared to those from 
gluino or chargino interactions, thus does not make much impact.

Larger asymmetries exceeding $10\%$ is attainable 
if one allows for only LL mixing, since the presence of 
RR mixing generates non-zero values for $c_{7}^{\prime }$, which
contributes to the branching ratio but not to $A_{{\rm CP}}$.
For example, for $(\widetilde{m}_{L,2},\ \widetilde{m}_{L,3})
=(100,\, 300)$ GeV, $A_{{\rm CP}}$ can reach 15\%. 

It is of interest to note that, in the case with 
LR mixing only, the SUSY contribution has 
a large enhancement factor $m_{\tilde{g}}/m_{b}$. 
To satisfy the bound from observed branching ratio, 
the squark masses need to be nearly degenerate if 
the mixing angles are not small. 
Furthermore, because the chargino 
contribution is small as mentioned before,
in this case it does not cancel against
the gluino contribution. 
In Fig. 2, we show $A_{\rm CP}$ for LR mixing with $u=v$, 
and with $\widetilde{m}=300$ GeV, $\Delta \widetilde{m}_{2}^{2}=
(20$ GeV)$^{2}$ and $\Delta \widetilde{m}_{3}^{2}=(30$ GeV)$^{2}$. 
We see that $A_{\rm CP}$ can reach 10\%.
Thus, even if down squark masses are large and nearly degenerate 
(i.e. near universal squark masses),
just some slight LR mixing could cause sizable $A_{\rm CP}$.

The B factories which would turn on soon will provide useful 
information about direct CP violation and can 
test the different models discussed here.

\section{Mixing Induced CP Violation}

For radiative $B_{d(s)}\to M_{d(s)}\gamma $ decays, 
where $M_{d(s)}$ is a $S=-1$($0$) CP eigenstate with 
eigenvalue $\xi =\pm $, it is possible to observe 
mixing induced CP violation \cite{Atwood}. 
Let $\Gamma (t)$ and $\bar{\Gamma}(t)$ be the time dependent rate 
for $B^0\to M^0\gamma $ and $\bar{B}^0 \to M^0\gamma $, respectively.
One has 
\begin{eqnarray}
R_{{\rm CP}} =
{\frac{\Gamma (t)-\bar{\Gamma}(t)}{\Gamma (t)+\bar{\Gamma}(t)}}
= -A_{{\rm mix}}\,\sin (\Delta m\, t), \ \ \
A_{{\rm mix}} =
\frac{2|c_{7}c_{7}^{\prime }|}{|c_{7}|^{2}+|c_{7}^{\prime}|^{2}}\,
\xi \,\sin [\phi _{B}-\phi -\phi ^{\prime }],
\end{eqnarray}
where $\Delta m$ and $\phi _{B}$ are the mass difference and phase 
in $B_{d(s)}-\bar{B}_{d(s)}$ mixing amplitude, 
and $\phi^{(\prime )}$ is the weak phase of $c_{7}^{(\prime)}$. 
In the $B_{d}$ case $\phi _{B}$ is the same as in SM since 
we do not consider squark mixings involving first generation.

In SM, $c_{7}^{\prime }/c_{7}=m_{s}/m_{b}$ hence 
$A_{{\rm mix}}^{{\rm SM}}$ is rather small. 
To obtain large $A_{{\rm mix}}$, 
both $c_{7}$ and $c_{7}^{\prime }$ have to be simultaneously sizable. 
This can be easily achieved in SUSY models. 
In Figs. 3 and 4, we show some representative results using 
the same parameters as in Figs. 1 and 2. We find that in these cases, 
$\sin 2\vartheta_{\rm mix}\equiv 
2|c_{7}c_{7}^{\prime}|/(|c_{7}|^{2}+|c_{7}^{\prime }|^{2})$ 
can reach 80\%, 90\%, respectively. 
It is interesting to note that 
large mixing induced CP violation ($A_{\rm mix}$)
does not necessarily imply large direct CP violation ($A_{\rm CP}$),
and vice versa.
In Fig. 5 we show the results in the LL=RR mixing case with 
$\widetilde{m}_{L,2}=\widetilde{m}_{R,2}=100$ GeV and 
$\widetilde{m}_{L,3}=\widetilde{m}_{R,3}=150$ GeV, 
where $\sin 2\vartheta_{\rm mix}$ can reach 80\%. 
We do not have large direct $A_{\rm CP}$ in this case, 
but one still can have large mixing induced CP violation.
We note that the allowed region is rather large. 
However, the constraint from $Br(B\rightarrow X_{s}\gamma)$ 
does not favor large mass spliting in LL=RR case. 
For example, it does not allow the choice 
$\widetilde{m}_{L,2}=\widetilde{m}_{R,2}=100$ GeV 
and $\widetilde{m}_{L,3}=\widetilde{m}_{R,3}=250$ GeV. 

To have large $A_{{\rm mix}}$, the phase combination 
$P=\sin (\phi _{B}-\phi -\phi ^{\prime })$ also needs to be large. 
This is easily achieved for LL and RR mixing case 
because the phase $\phi ^{\prime }$ is not constrained by
the observed branching ratio. 
In the case with LR mixing only, because of the assumption of $u=v$, 
the phases are related and therefore are constrained from 
the interference with SM contribution in the branching ratio. 
One needs to make sure that the factor $P$ is also large. 
We have checked in detail that this indeed happens 
in the cases considered. 
One can also relax the requirement to allow $c_{7}^{\prime }$ 
to have an independent phase. 
In this case the factor $P$ can always be made large. 
Note that, even if $\phi $ and $\phi ^{\prime }$ vanish 
(no CP violation from soft squark masses), nonvanishing $\phi _{B}$ 
from SM contribution to $B$--$\bar B$ mixing can still lead to 
observable $A_{{\rm mix}}$, 
so long that $c_{7}$ and $c_{7}^{\prime }$ are comparable.

We have only assumed that the state $M$ be a CP eigenstate, 
which can be $K_S\pi^0$ from $K^{*0}$ or $K^{*0}_{1,2}$ 
for $B_d$ decays, or $\phi$ for $B_s$ decay. 
The expression for $A_{{\rm mix}}$ is process independent. 
However, because of the relatively long lifetime of $K_S$, 
and the fact that having $\gamma$ and $\pi^0$ in the final state 
do not provie good determination of the decay vertex position, 
$A_{{\rm mix}}$ for $B_d \to K^{*0} \gamma$ probably can not be 
measured with sufficient accuracy. 
Perhaps the $B_d \to K^*_{1,2} \gamma$ situation would be better, 
but {\it these modes have to be measured first}. 
The situation for $B_s\to \phi \gamma$ is definitely better, 
but it can only be carried out at hadronic facilities such as 
the Tevatron or LHC, and only after $B_s$ mixing is measured.

\section{$\Lambda$ Polarization in Beauty Baryon Decay}

The chiral structure can be easily studied in 
$\Lambda _{b}\to \Lambda\gamma $. The decay amplitude is given by 
\begin{equation}
A(\Lambda _{b}\to \Lambda \gamma )
=-{\frac{G_{F}}{\sqrt{2}}}{\frac{e}{8\pi^{2}}}V_{tb}V_{ts}^{*}\,
C\,\bar{\Lambda}[c_{7}(1+\gamma _{5})+c_{7}^{\prime}(1-\gamma _{5})]
\sigma_{\mu \nu }F^{\mu \nu }\Lambda _{b},
\end{equation}
where $C$ is a form factor which can in principle be 
calculated in heavy quark effective theory. 
The resulting branching ratio is of order $10^{-5}$
and should be measurable at future hadronic B factories. 
The chiral structure can be studied by measuring the polarization 
of $\Lambda $, via the angular distribution \cite{Mannel} 
\begin{equation}
\frac{1}{\Gamma }{\frac{d\Gamma }{d\cos \theta }}
={\frac{1}{2}}\,(1+\alpha_{\Lambda }\cos \theta ),\ \ \ \ \ \ \ 
\alpha _{\Lambda }={\frac{|c_{7}|^{2}-|c_{7}^{\prime }|^{2}}
{|c_{7}|^{2}+|c_{7}^{\prime }|^{2}}},
\label{baryon}
\end{equation}
where $\theta $ is the angle between the 
direction of the momentum of $\Lambda $ in 
the rest frame of $\Lambda _{b}$ and the 
direction of the $\Lambda $ polarization in its rest frame. 
We emphasize that the parameter $\alpha _{\Lambda }$ 
does not depend on the hadronic parameter $C$, which
makes it a good quantity for studying the chiral structure 
without any uncertainties from hadronic matrix elements. 
In SM one has $\alpha _{\Lambda}=1$. 
Deviation from this value for $\alpha _{\Lambda }$ would be an
indication of physics beyond SM. 
Since Eq. (\ref{baryon}) does not depend on
the hadronic matrix element of the specific process, 
it can be applied to any other radiative beauty baryon decays.

It is clear that if $c_{7}$ and $c_{7}^{\prime }$ are 
of the same order, one would have substantial deviation 
from SM prediction. But unlike the case for $A_{{\rm mix}}$, 
large $\alpha _{\Lambda }$ does {\it not} require a large
phase combination factor $P$. 
In fact, $\alpha _{\Lambda }$ is a measure of
the chiral structure independent of CP violation. 
One can of course still study direct CP violation rate asymmetries. 
If the new physics contribution comes only from LL mixing, 
one would not have large deviation from SM prediction. 
In Figs. 6,  7 and 8, we show the deviation from SM prediction, 
$1-\alpha_{\Lambda }$, for the cases in Figs. 1, 2 and 5. 
We see that $\alpha_{\Lambda }$ can indeed be very different from SM. 
Note that in these cases $\alpha _{\Lambda }$ has the same sign 
as in SM. 
For LL/RR mixing cases, this has to do with 
the compensating effect between chargino and gluino loops,
while for LR mixing case it has to do with the 
heaviness of squarks.
We have only explored a small portion of the parameter space
as a consequence of our many simplifications.
Although we have not been able to identify parameter space where 
$\alpha_\Lambda$ flips sign,
it does not mean that this is impossible.

\section{Discussion and Conclusions}

In this paper we have shown that SUSY models with 
nonuniversal squark masses can give rise to 
rich phenomena in $b\to s\gamma$ decays. 
Indeed, such considerations received attention with 
the notion that $|c_8 | \sim 2$ could help resolve \cite{bsg,susy} 
the long-standing low charm counting and semileptonic 
branching ratio problems, in the form of a rather enhanced 
$b\to sg \sim 5\%$--10\%. It has been shown that this is 
possible in SUSY models \cite{susy}, 
but $b\to s\gamma$ provides a severe constraint.
However, to have large SUSY effect in $b\to s\gamma$ decay, 
$b\to sg$ need not be greatly enhanced. 
We have incorporated into the model the consideration of 
new CP phases, which naturally arise. 
Although we do not claim to have explored the
full parameter space, we find that SUSY with 
nonuniversal squark masses could indeed lead to dramatic effects. 
The severe constraint from $Br(B\rightarrow X_{s}\gamma )$ 
does not exclude squark mixings. We find a 
cancellation effect between gluino and chargino contributions, 
which gives rise to rather large allowed regions for 
the mixing angle $\theta $ and phase $\sigma $, 
leading to interesting consequences for CP violation.
The features exhibited in the present analysis is 
a common feature in SUSY models with low energy flavor 
and CP violating squark mixings.
A model with mixing only between second and third generation
down type squarks can easily evade known low energy constraints
but give dramatic signals in $b\to s$ transitions.

Our purpose has been to illustrate such efficacy and hopefully 
motivate our experimental colleagues for detailed studies. 
We find that direct CP violating rate asymmetries can be 
as large as 10\%, comparable to 
general multi-Higgs doublet models\cite{Wolfenstein}. 
Purely LL mixing is favored in this case. 
Even more interesting would be the observation of 
mixing induced CP violation. Here, 
purely LL or RR mixing is insufficient, but LL and RR mixing 
or LR mixing models could lead to rather sizable effects. 
The observation of mixing induced CP violation immediately 
demonstrates that $b\to s\gamma $ decay has two chiralities. 
Independent of CP violation, however, the chirality structure 
can be tested by studying $\Lambda$ polarization in 
$\Lambda _{b}\to \Lambda \gamma $ decay. 
The parameter $\alpha _{\Lambda }$ can deviate from the SM value of 1. 
The nonobservance of CP violation in $B$ meson decays 
does not preclude surprises in the $\alpha_{\Lambda }$ measurement.

It is clear that $b\to s\gamma $ transitions provide good tests for new
physics.

This work is supported in part by grant NSC 87-2112-M-002-037, NSC
87-2811-M-002-046, and NSC 88-2112-M-002-041 
of the Republic of China, 
and by Australian Research Council.
We wish to thank the referee for urging us to include chargino effects,
which turned out to be rather significant.


\begin{figure}[htb]
\centerline{\DESepsf(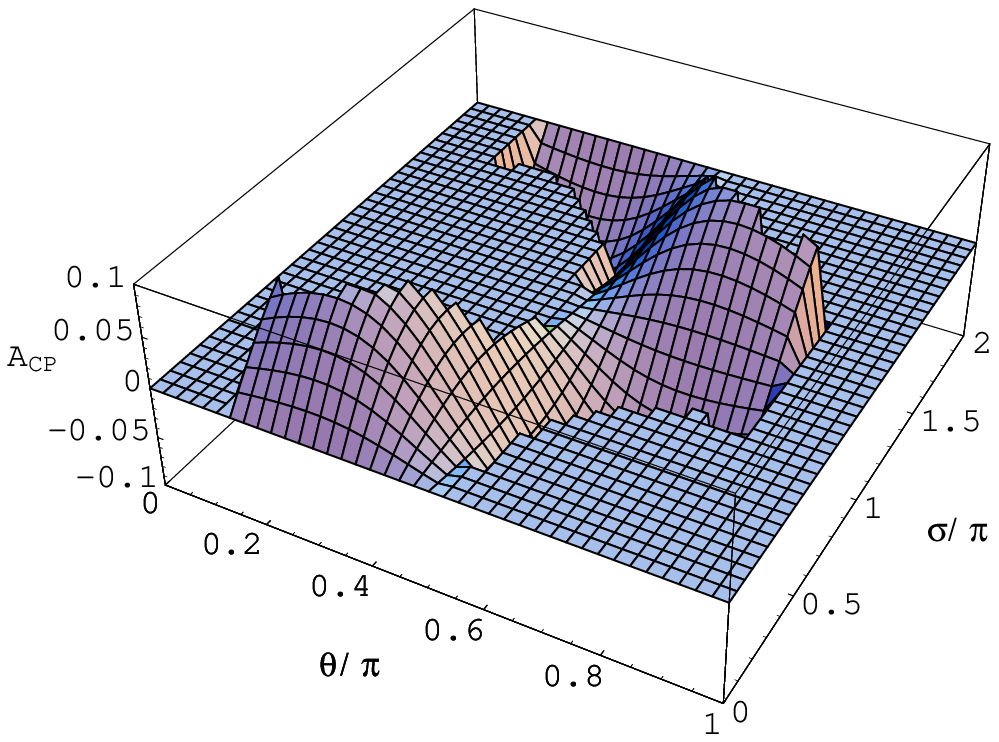 width 10cm)}
\smallskip
\caption{$A_{\rm CP}$ vs. mixing angle $\theta$ and phase $\sigma$ 
for LL and RR but no LR mixings, with the mass values 
$m_{\tilde{g}}=200$ GeV and $(\widetilde m_{L,2}$, 
$\widetilde m_{L,3}$, $\widetilde m_{R,2}$, 
$\widetilde m_{R,3})= (100$, 250, 100, 150) GeV. 
The flat surface corresponding to $A_{\rm CP}= 0$ 
is the parameter space
forbidden by the $B\rightarrow \, X_s \gamma$ constraint, and the cut on 
$\theta$ is due to lower bound of stop mass.}
\end{figure}

\begin{figure}[htb]
\centerline{\DESepsf(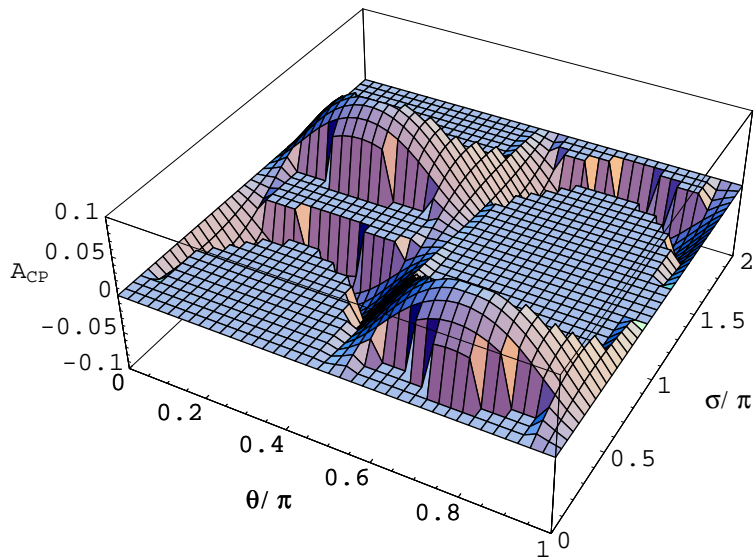 width 10cm)}
\smallskip
\caption{
$A_{\rm CP}$ for LR mixing with $u=v$ and $M_{\tilde{g}}=200$ GeV,
$\widetilde m=300$ GeV, $\Delta \widetilde m_{2}^2=(20$ GeV)$^2$ and 
$\Delta \widetilde m_{3}^2= (30$ GeV)$^2$. 
}

\end{figure}

\begin{figure}[htb]
\centerline{\DESepsf(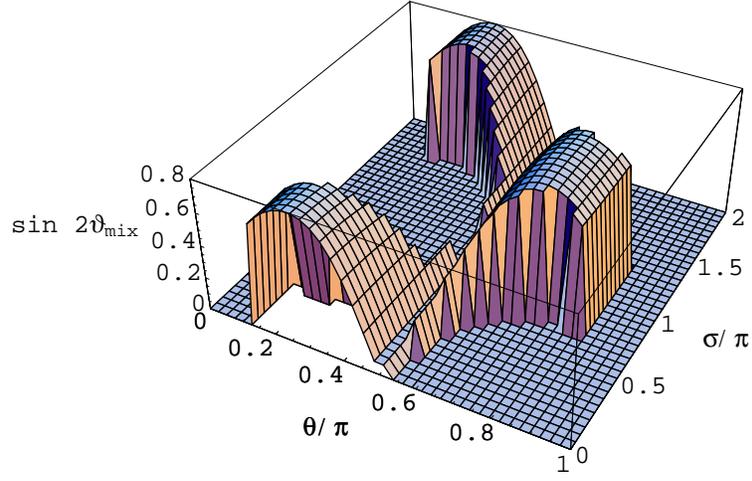 width 10cm)}
\smallskip
\caption{$\sin (2\,\vartheta_{mix})$ in LL and RR mixing
case with same parameter space as in Fig. 1.}
\end{figure}

\begin{figure}[htb]
\centerline{\DESepsf(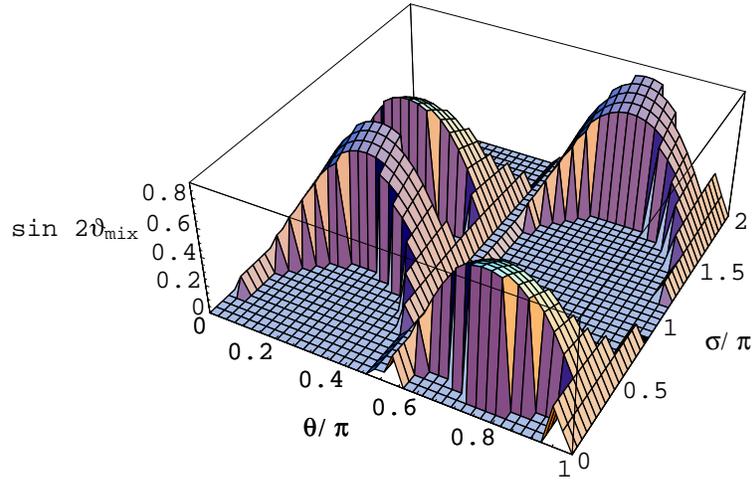 width 10cm)}
\smallskip
\caption{$\sin (2\,\vartheta_{mix})$ in LR mixing
case with same parameter space as in Fig. 2.}
\end{figure}

\begin{figure}[htb]
\centerline{\DESepsf(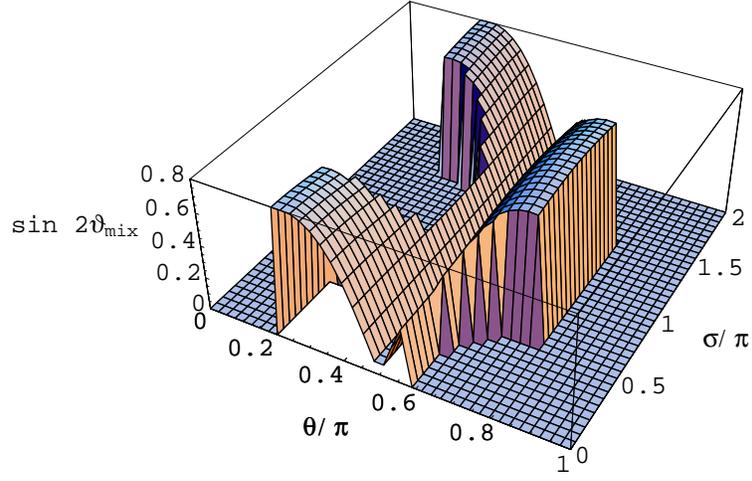 width 10cm)}
\smallskip
\caption{$\sin (2\,\theta_{mix})$ in LL=RR mixing case with 
$M_{\tilde{g}}=200$ GeV, $\widetilde m_{L,2}=\widetilde m_{R,2}=100$ GeV
and $\widetilde m_{L,3}=\widetilde m_{R,3}=150$ GeV.} 
\end{figure}

\begin{figure}[htb]
\centerline{\DESepsf(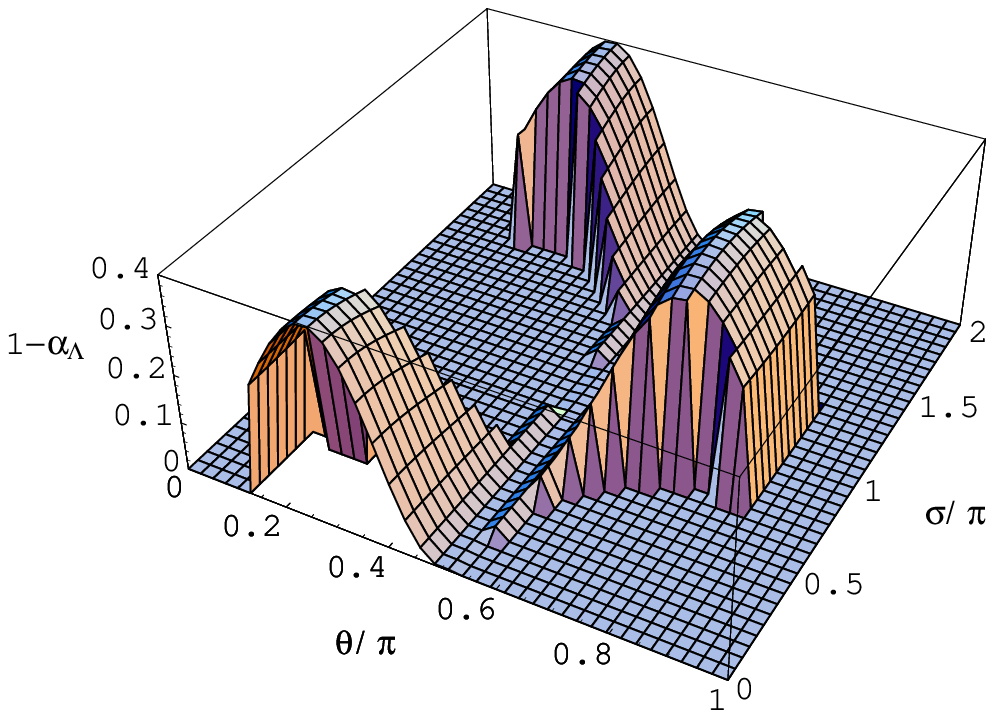 width 10cm)}
\smallskip
\caption{$1-\alpha_{\Lambda}$ in LL and RR mixing
case with same parameter space as in Fig. 1.}
\end{figure}

\begin{figure}[htb]
\centerline{\DESepsf(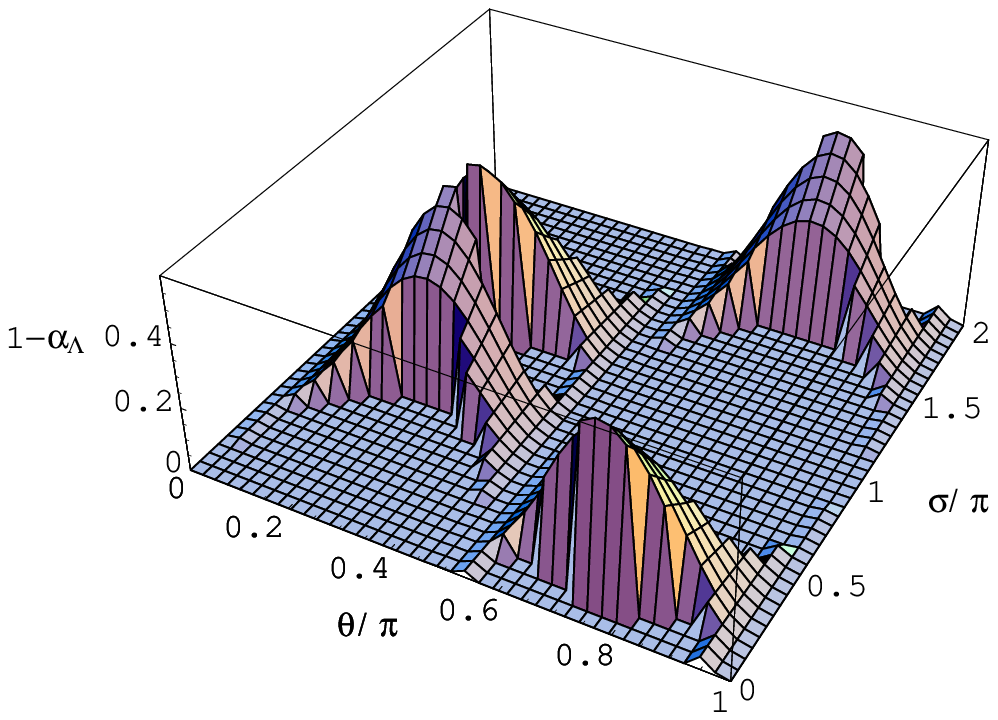 width 10cm)}
\smallskip
\caption{$1-\alpha_{\Lambda}$ in LR mixing
case with same parameter space as in Fig. 2.}
\end{figure}

\begin{figure}[htb]
\centerline{\DESepsf(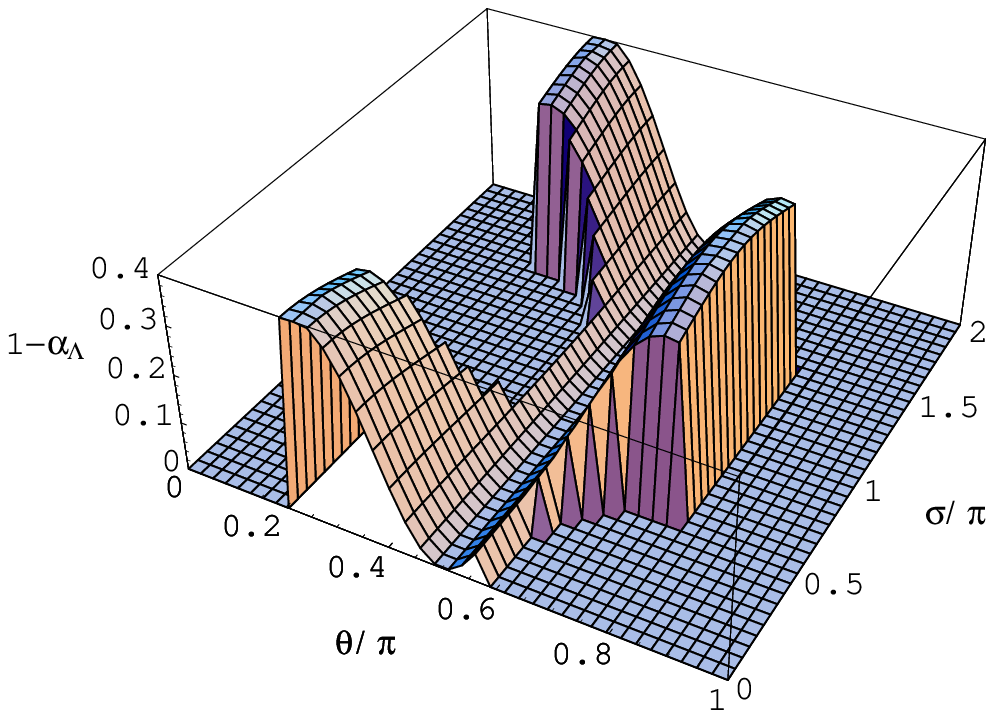 width 10cm)}
\smallskip
\caption{$1-\alpha_{\Lambda}$ in LL=RR mixing
case  with same parameter space as in Fig. 5.}
\end{figure}

\end{document}